\documentclass[onecollarge,natbib]{svjour2}
\bibpunct{[}{]}{,}{n}{}{,} 
\smartqed  
\usepackage{graphicx}
\usepackage{latexsym}
\usepackage{marvosym}
\journalname{Few-Body Systems (APFB2011)}
\begin{document}

\title{\boldmath
On the Mass Difference between $\pi$ and $\rho$ 
\\ using a Relativistic Two-Body Model
}


\author{Jin-Hee Yoon \and Byeong-Noh Kim \and \\ Horace W. Crater
        \and Cheuk-Yin Wong}


\institute{
	Jin-Hee Yoon(\Letter) and Byeong-Noh Kim  \at 
            Department of Physics, Inha University, Incheon
            402-751, Republic of Korea \\
           \email{jinyoon@inha.ac.kr}           
     \and
	Horace W. Crater \at The University of Tennessee Space Institute, 
	        Tullahoma, Tennessee 37388, USA \\
     \and
	Cheuk-Yin Wong \at Physics Division, Oak Ridge National laboratory, 
			Oak Ridge, Tennessee 37831, USA \\
}

\date{Received: date / Accepted: date}

\maketitle

\begin{abstract}
The big mass difference between the pion($\pi$) and rho meson($\rho$) 
possibly originates from the spin-dependent nature of the interactions in the two states since
these two states are similar except for spin.
Both $\pi$ and $\rho$ are quark-antiquark systems which can be treated using the two-body Dirac equations (TBDE) of constraint dynamics. 
This relativistic approach for two-body system has the advantage over the non-relativistic treatment in the sense that the spin-dependent nature is automatically coming out from the formalism. 
We employed Dirac's relativistic constraint dynamics to describe quark-antiquark systems. 
Within this formalism, the 16-component Dirac equation is reduced to the 4-component 2nd-order differential equation and the radial part of this equation is simply a Schr\"{o}dinger-type equation with various terms calculated from the basic radial potential. 

We used a modified Richardson potential for quark-antiquark systems which satisfies the conditions of confinement and asymptotic freedom. We obtained the wave functions for these two mesons which are not singular at short distances. We also found that the cancellation between the Darwin and spin-spin interaction terms occurs in the $\pi$ mass but not in the $\rho$ mass and this is the main source of the big difference in the two meson masses.

\keywords{pion mass \and two body dirac equation \and quark-antiquark bound state}
\end{abstract}

\section{Introduction}
\label{intro}

The $\pi$ and $\rho$ are similar states consisting of an $u$-quark and a $\bar{d}$-antiquark except for the spins as $\pi$ is in the spin-singlet state while $\rho$ is in the triplet. 
Therefore we can easily attribute any differences to the spin-dependent nature of the interaction in the two states.
However their mass difference is quite large in comparison with the mass differences between other meson pairs such as $\eta_C$ and $J/\Psi$ ($c\bar{c}$) or $D_0$ and $D^*_0$ ($c\bar{u}$). 
In this ground the large mass difference might also be due to the small mass of $\pi$, which is explained as a result of the chiral symmetry breaking in QCD. 
Therefore it will be helpful to identify the origin of the small mass of $\pi$ and to trace the big mass difference between $\pi$ and $\rho$.

Both $\pi$ and $\rho$ are quark-antiquark systems which can be described within non-relativistic framework.
However quarks are light particles and interact strongly; they should be treated relativistically. 
Furthermore, spin-dependence is purely relativistic in nature and cannot be derived non-relativistically. 
Therefore we need a relativistic description rather than a non-relativistic description, for the quark-antiquark systems. 
Todorov has suggested the two-body relativistic equations for spinless two-body systems \cite{todorov}.
Based on the Dirac's constraint formalism and a minimal interaction structure the TBDE has been tested and confirmed in classical \cite{crater92} and quantum field theory \cite{jallouli97}. 
Within this formalism, two coupled 16-component Dirac equation with 2 fermion basis is reduced to the single 4-component 2nd-order differential equation and the radial part of this equation are simply Schr\"{o}dinger-type with various terms calculated from a basic radial potential, which can be phenomenologically adopted. 
We have constructed the two-body Dirac equations (TBDE) \cite{alstine82,yoon09} for two interacting fermions in a form that is as easy to handle as the non-relativistic Schr\"{o}dinger equation. 
The brief description of this formalism is summarized in Sec. 2.
For the interactions between a quark and an antiquark we adopt some phenomenological potential, which will be described in Sec. 3.
Final numerical results and discussions will be presented in Sec. 4.

\section{The Relativistic Schr\"{o}dinger-like equation}
\label{sec:2}
Van Alstine and Crater utilized the two body Dirac constraint dynamics for two interacting fermions with mass-shell constraint to reduce the relative energy and time degrees of freedom \cite{crater92,alstine82}.
They applied the Pauli reduction and scale transformation and succeeded in reducing the 16 component Dirac equation into a 4 component relativistic Schr\"{o}dinger-type equation,
\begin{equation}
\label{schreq} 
{\cal{H}} \psi = {{\epsilon_1 {\cal H}_1 + \epsilon_2 {\cal H}_2} \over w} \psi
= \left\{ p^2 + \Phi_w (\sigma_1, \sigma_2, p_\perp, A(r), S(r)) \right\} \psi
=b^2(w) \psi,
\end{equation} 
where $w$ is the invariant mass and $b^2(w)=\epsilon^2_w-m^2_w$ with the relativistic energy $\epsilon_w$ and the relativistic reduced mass $m_w$.
$\Phi_w$ plays a role of interaction in the Schr\"{o}dinger-type equation (\ref{schreq}) and includes many relativistic terms: Darwin($D$), spin-spin($SS$), tensor($T$), spin-orbit($SO$), etc.
For a scalar potential $S$ and a time-like vector potential $A$,
$\Phi_w$ can be expressed as follows \cite{yoon09} :
\begin{eqnarray} 
\label{interaction}
\Phi_w &=& 2m_wS+S^2+2\epsilon_wA-A^2+\Phi_D+\L\cdot(\sigma_1+\sigma_2)\Phi_{SO}
+\sigma_1\cdot\sigma_2\Phi_{SS} \\
\nonumber
&&+\sigma_1\cdot\hat{r}\sigma_2\cdot\hat{r} \L\cdot(\sigma_1+\sigma_2)\Phi_{SOT}
+(3\sigma_1\cdot\hat{r}\sigma_2\cdot\hat{r}-\sigma_1\cdot\sigma_2)\Phi_{T} \\
\nonumber
&&+\L\cdot(\sigma_1-\sigma_2)\Phi_{SOD}
+i\L\cdot(\sigma_1\times\sigma_2)\Phi_{SOX}.
\end{eqnarray} 

If $S$ and $A$ are central radial potentials, this equation is reduced further into two-coupled radial equations.
For $\pi$ which is in a spin-singlet state, the differential equation becomes 
\begin{equation}
\label{pieq}
\left\{ -{d^2\over dr^2}+2m_wS+S^2+2\epsilon_wA-A^2+\Phi_D-3\Phi_{SS}
\right\} v_0=b^2v_0.
\end{equation} 
$\rho$ is in a spin-triplet state, which is a superposition of the S- and D-states. 
These two states are coupled by the following differential equations 
\begin{equation} 
\label{rhoeq1}
\left\{ -{d^2\over dr^2}+2m_wS+S^2+2\epsilon_wA-A^2+\Phi_D+\Phi_{SS}
\right\} u_S 
+{2\sqrt{2} \over 3}(3\Phi_T-6\Phi_{SOT})u_D =b^2 u_{S}, \\
\end{equation}
\begin{eqnarray}
\label{rhoeq2}
\left\{ -{d^2\over dr^2}+{6\over r^2}+2m_wS+S^2+2\epsilon_wA-A^2+\Phi_D-6\Phi_{SO}+\Phi_{SS}-2\Phi_{T}+2\Phi_{SOT}
\right\} u_D \\ \nonumber
+{2\sqrt{2} \over 3}(3\Phi_T)u_S =b^2 u_{D}.
\end{eqnarray} 
Here $v_0=\psi_\pi(r)/r$, and $u_S$ and $u_D$ are corresponding quantities of $\rho$ for the S-state and D-state, respectively. 
Solving these equations are not easy but at least manageable numerically.

\section{Phenomenological Potential}
\label{sec:3}
To complete the differential equation, we need a basic potentials between a quark and an antiquark. 
Commonly used potential for two quarks are the static Cornell potential, 
which is composed of a static Coulomb-type potential and a linear potential \cite{eichten80}.
However this potential does not reflect the asymptotic freedom which is one of the important characteristics of the $q-\bar{q}$ interaction. Richardson\cite{richardson79} has suggested the following form for heavy quarkonia,
\begin{equation} 
\tilde{V}(q) = -{ 16\pi\over 27}{1 \over {q^2 \ln (1+q^2 /\Lambda^2)}}
\end{equation} 
in momentum space.
After Fourier transform to coordinate system, it becomes 
\begin{equation} 
V(r) = { 8\pi\Lambda^2r \over 27}-{ 8\pi f(\Lambda r) \over 27r}.
\end{equation} 
The function $f(\Lambda r)$ needs to approach $-1/\ln (\Lambda r)$ as $r \rightarrow 0$ and to approach 1 as $r \rightarrow \infty$ for the potential satisfying both crucial characteristics in QCD, the asymptotic freedom and the confinement.
We modified this potential using 3 parameters, $\Lambda$, $K$, and $B$,
\begin{equation} 
\label{modRpot}
V(r) = { 8\pi\Lambda^2r \over 27}-{ 16\pi \over 27r \ln (Ke^2+B/(\Lambda r)^2 )}.
\end{equation} 
These 3 parameters will be determined to fit the experimental values of meson masses.
The one feature of this method is that there is no cutoff parameter introduced.
Everything is treated non-perturbatively and it gives good results for heavy quark meson systems as well as light two quark systems.

Partitioning the confinement term in $V(r)$ of Eq.(\ref{modRpot}) as an invariant $S$ responsible for the scalar potential and the Coulomb-like term as an invariant $A$ responsible for the full four-vector potential, we can obtain the specific forms for the various terms in Eq.(\ref{interaction}). The readers can refer the Eqs.(A17)-(A20) in the Ref.\cite{yoon09} for their explicit forms.

\section{Results and Conclusion}
\label{sec:3}

\begin{figure}[b]
\centering
\includegraphics[width=0.55\textwidth]{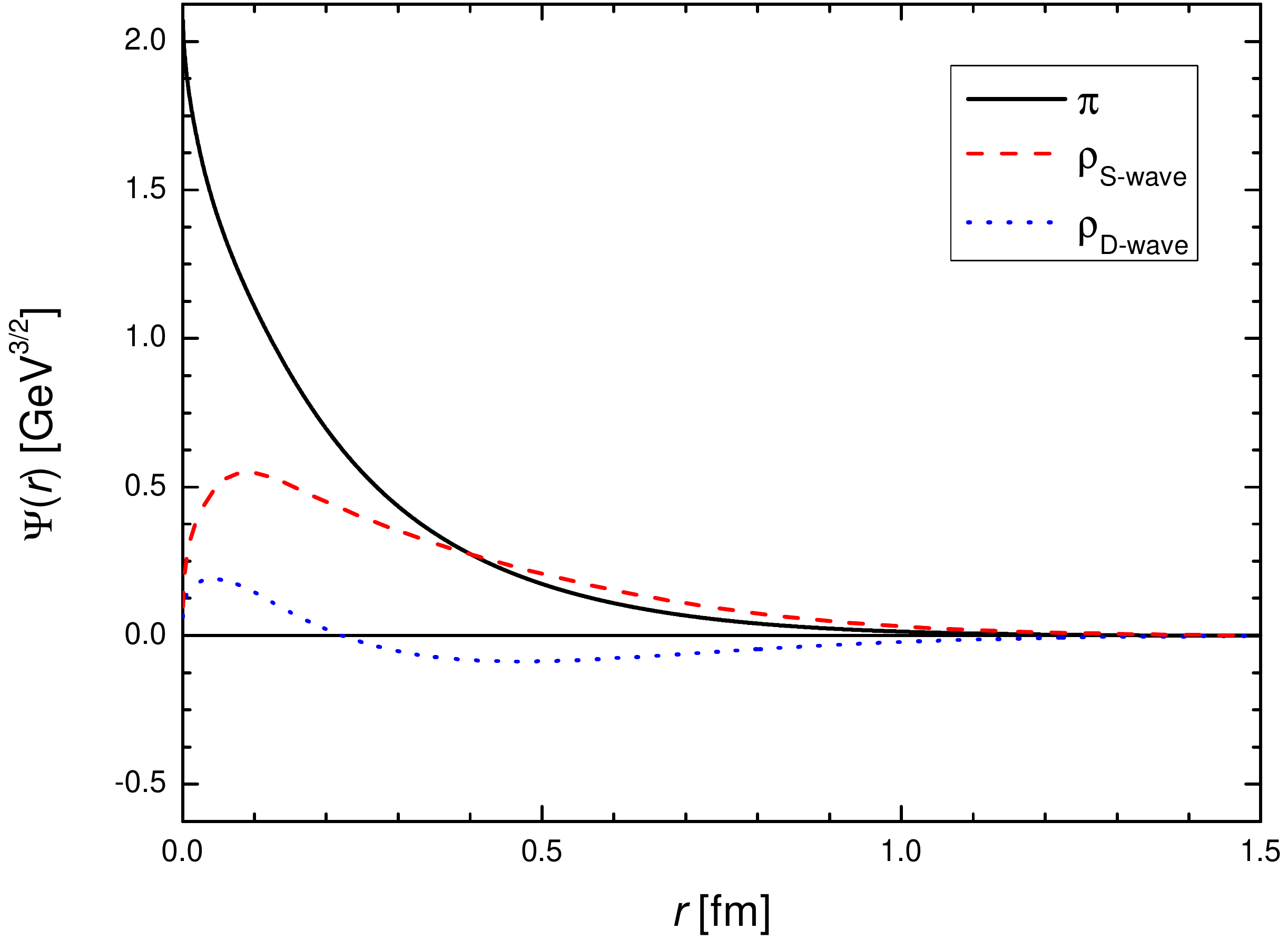}
\caption{The wave functions of $\pi$ and $\rho$. 
The solid (black in color) line is $\pi$ wave function. 
The dashed (red in color) curve is the S-state wave function of $\rho$ and 
the dotted (blue in color) curve is the D-state wave function of $\rho$.}
\label{fig:1}
\end{figure}
We fit the invariant masses of 32 mesons and determine 3 parameters in the potential. Their values are $\Lambda=0.4218$ GeV, $B=0.05081$, $K=4.198$. 
Here the quark masses used to give the best fit are $m_u=0.0557$ GeV, $m_d=0.0553$ GeV. 
Then using these values we solve the differential equations, Eq.(\ref{pieq}) for $\pi$ and Eqs.(\ref{rhoeq1}) and (\ref{rhoeq2}) for $\rho$,
and obtain their wave functions. 
They are shown in Fig.1.
The wave functions for both mesons are not singular at short distances and the asymptotic behavior of D-state wave function of $\rho$ is proportional to that of its S-state wave function at small radii.

With the given wave functions we can obtain the eigenvalues, $b^2$, which in turn give the invariant masses of 0.159 GeV for $\pi$ and 0.792 GeV for $\rho$, while their experimental values are 0.140 GeV and 0.775 GeV, respectively.
We tabulated the contribution of each term in Eqs.(\ref{pieq})-(\ref{rhoeq2}) in Table \ref{tab:0}.
The spin-spin contribution ($<-3\Phi_{SS}>$) for pion is larger than the other spin-dependent terms by an order of magnitude. 
However for $\rho$ the same contribution ($<\Phi_{SS}>$) is smaller than the other spin-dependent contributions by an order of magnitude. 
Therefore the spin-dependent term plays crucial role for the mass difference between $\pi$ and $\rho$. 
The major contribution for pion mass comes from the Darwin ($\Phi_{D}$) and spin-spin interaction terms. The magnitudes of these two terms are about same but opposite in sign, and therefore they almost cancel each other, resulting in a small pion mass. 
However for the $\rho$, the Darwin term is not dominant at all. 
It is about the same as or less than the off-diagonal contributions from tensor ($\Phi_{T}$) or spin-orbit-tensor ($\Phi_{SOT}$) terms. 
Besides that the spin-spin contribution is much smaller than the Darwin or the 
off-diagonal terms, it contributes in same direction as the Darwin term, as distinct from the pion case. 
Overall the $\rho$ mass is not that small. 
We can also notice that the kinetic energy 
contributions ($<-d^2/dr^2 + J(J+1)/r^2>$) are
almost 1 GeV for $\pi$ and 0.5 GeV for $\rho$. 
These numbers are much larger than the quark masses of $\sim 0.0557$ GeV and
it justifies to treat the quarks relativistically.

Finally we can conclude that the cancellation between the Darwin and spin-spin interaction terms occurs in $\pi$ mass but not in $\rho$ mass and this is the main source of a big difference in two masses.
Also we can say that our theory is consistent with the chiral symmetry breaking which causes the non-zero small mass of pion.

\begin{table}[t]
\caption{The contributions of all terms in Eqs.(\ref{pieq})-(\ref{rhoeq2}). 
         The column $<00>$ is for $\pi$ with S-state only 
         and the remaining four columns are the diagonal and off-diagonal terms
         for $\rho$ with both S- and D-state wave functions.
         The numbers in () are experimental values. 
         All terms except $w$ are in a unit of $\mbox{GeV}^2$ and $w$ is in 
         a unit of GeV. }
\centering
\label{tab:0}       
\begin{tabular}{c||c||r|r|r|r}
\hline\noalign{\smallskip}
& $\pi$ & \multicolumn{4}{c}{$\rho$} \\
\cline{2-6}
Terms & $<00>$ & $<SS>$ & $<DD>$ & $<SD>$ & $<DS>$ \\
 \tableheadseprule \hline
$<-d^2/dr^2 + J(J+1)/r^2>$ & {\bf 0.8508} & 
                          {\bf 0.3085} & {\bf 0.2812} & &  \\
 \hline 
$\matrix{<2m_w S> \cr <S^2> \cr <2 \epsilon_w A> \cr <-A^2 > \cr 
{\bf \mbox{Sub Total}}}$ & 
$\matrix{~~0.0103 \cr ~~0.0942 \cr -0.0598 \cr -0.4279 \cr {\bf -0.3832} }$ & 
$\matrix{~~0.0026 \cr ~~0.1631 \cr -0.2091 \cr -0.1247 \cr {\bf -0.1680}}$ & 
$\matrix{~~0.0006 \cr ~~0.0446 \cr -0.0211 \cr -0.0094
         \cr ~~{\bf 0.0146}}$ & &
\\ \hline
$\matrix{<\Phi_D> \cr <-3\Phi_{SS}> \cr <\Phi_{SS}> \cr <-2\Phi_T> 
         \cr <-6\Phi_{SO}> \cr <2\Phi_{SOT}> \cr {\bf \mbox{Sub Total}}}$ & 
$\matrix{-3.804 \cr ~~3.340 \cr\cr\cr\cr\cr {\bf -0.4643}}$ &  
$\matrix{ -0.2790 \cr\cr -0.0364 \cr  \cr \cr \cr {\bf -0.3154}}$ &
$\matrix{ -0.0436 \cr\cr -0.0095 \cr -0.0417 \cr ~~0.1133 \cr ~~0.0215 
         \cr ~{\bf 0.0346}}$ & & \\
 \hline 
$\matrix{(2\sqrt{2}/3)<3\Phi_T> \cr (2\sqrt{2}/3)<-6\Phi_{SOT}> 
         \cr {\bf \mbox{Sub Total}}}$ & & & &
$\matrix{ -0.3088 \cr ~~0.6177 \cr ~~{\bf 0.3090}}$ &
$\matrix{ -0.3088 \cr \cr {\bf -0.3088}}$ \\
 \tableheadseprule \hline 
Total Sum & {\bf 0.0033} & {\bf -0.1750} & {\bf 0.3158} & {\bf 0.3090} 
          & {\bf -0.3088} \\
\hline
$b^2$ & {\bf 0.0033} & \multicolumn{4}{c}{ {\bf 0.1411}} \\
\hline
$w(\mbox{GeV})$ & {\bf 0.159 (0.140)} & \multicolumn{4}{c}{~ {\bf 0.792 (0.775)}} \\
\hline
\end{tabular}
\end{table}

\vspace{-0.3cm}
\begin{acknowledgements}
This work was supported by National Research Foundation of Korea under the program number 2011-0003707 and by the Office of Nuclear Physics, U.S. Department of Energy. 
\end{acknowledgements}

\vspace{-0.6cm}



\end{document}